%% file: main.tex
\documentclass[a4paper]{jpconf}
\usepackage{graphicx}
\usepackage{subfig}

\usepackage{lineno}

\graphicspath{{figs/}}
\input{definitions.tex}

\begin{document}
\title{Fast Data-Driven Simulation of Cherenkov Detectors Using Generative Adversarial Networks}

\author{A~Maevskiy$^1$, D~Derkach$^1$, N~Kazeev$^{1,2,3}$, A~Ustyuzhanin$^{1,2}$, M~Artemev$^1$ and L~Anderlini$^4$\\
on behalf of the LHCb collaboration}
\address{$^1$ Laboratory of Methods for Big Data Analysis, National Research University Higher School of Economics, 3~Kochnovsky Proezd, Moscow 125319, Russia}
\address{$^2$ The Yandex School of Data Analysis, 11/2~Timura Frunze St., Moscow 119021, Russia}
\address{$^3$ Department of Physics, Sapienza University of Rome,  5 Piazzale Aldo Moro, Rome 00185, Italy}
\address{$^4$ Istituto Nazionale di Fisica Nucleare, Sezione di Firenze, via G. Sansone 1, Sesto Fiorentino 50019, Italy}

\ead{artem.maevskiy@cern.ch}

\begin{abstract}
    The increasing luminosities of future Large Hadron Collider runs and next generation of collider experiments will require an unprecedented amount of simulated events to be produced. Such large scale productions are extremely demanding in terms of computing resources. Thus new approaches to event generation and simulation of detector responses are needed. In LHCb, the accurate simulation of Cherenkov detectors takes a sizeable fraction of CPU time. An alternative approach is described here, when one generates high-level reconstructed observables using a generative neural network to bypass low level details. This network is trained to reproduce the particle species likelihood function values based on the track kinematic parameters and detector occupancy. The fast simulation is trained using real data samples collected by LHCb during run~2. We demonstrate that this approach provides high-fidelity results.
\end{abstract}

\section{Introduction}

Simulation of particle collisions occurring at the Large Hadron Collider (LHC) provides a detailed theoretical reference for the measurements performed at LHC experiments.
The demand on the number of simulated events is growing rapidly with the increase of luminosity at the LHC. Given the computational requirements of accurate detector simulation algorithms, it becomes unfeasible to use them to fulfill typical requests for the modelled events from physics analyses at LHC. Thus faster approaches to event generation and simulation are needed.

The LHCb detector~\cite{Alves:2008zz} is one of the four major experiments at the LHC in CERN. It is a single-arm forward spectrometer, designed to study particles containing $c$- and $b$-quarks, which requires robust particle identification (PID) system. PID in LHCb is provided by four different subsystems: the calorimeter system, the two Ring-imaging Cherenkov (RICH) detectors and the muon stations. Simulating the RICH detectors is particularly computationally heavy due to the need to accurately model the low-energy secondary electrons, as well as the light propagation, diffraction and absorption effects~\cite{Easo:2005xv}.

In this paper, we propose a novel solution to the problem of fast simulation for the RICH detectors at LHCb. The proposed solution bypasses accurate RICH simulation algorithms and uses the approach of generative adversarial networks (GANs)~\cite{Goodfellow:2014upx} to generate the high-level reconstructed observables.

\subsection{Generative adversarial networks}

The key idea behind GANs involves simultaneous training of two neural networks. One network, named \emph{generator}, takes samples from a known distribution (e.g. Gaussian, typically multivariate) and transforms them to the output that is trained to be distributed similarly to data. The other, \emph{discriminator}, is given both data and generator's output as input, and is trained to distinguish between the two.

The training of the two networks occurs in turns, and typically the loss of the generator is constructed to be the negative from that of the discriminator. Such adversarial learning setup can become unstable in case one network outperforms the other, and further regularization procedures and/or hyperparameter tuning might be needed to achieve convergence. In~\cite{Goodfellow:2014upx}, the metric optimized by the discriminator is cross entropy, which leads to overall equilibrium achieved when Jensen-Shannon (JS) divergence between the data and the generated samples is minimized. Other metrics, such as Wasserstein distance and Cramer distance, were proposed for GANs and have already proved to converge faster~\cite{WGAN, DBLP:journals/corr/GulrajaniAADC17, DBLP:journals/corr/BellemareDDMLHM17}. For example, Wasserstein and Cramer GANs have previously been used in particle and astroparticle
physics tasks~\cite{Derkach:2019qfk, Erdmann2018, Erdmann2019}. The main advantage of the Wasserstein and Cramer metrics over JS is that they provide a smooth measure even for disjoint distributions, they prevent mode collapses, when generator learns to cover only a part of the data distribution, and solve the problem of vanishing gradients for the case of discriminator outperforming the generator. In addition, Cramer metric avoids the problem of biased gradients~\cite{DBLP:journals/corr/BellemareDDMLHM17}, and therefore it is the metric of our choice.

\section{Overview of the data}

RICH detectors make use of the Cherenkov effect to identify particles. An ultrarelativistic particle traversing through a transparent medium emits Cherenkov photons within a cone whose spread angle is a function of particle's velocity. Therefore, measuring this angle and momentum can constrain the mass of the particle and thus provide the necessary PID information. In RICH, the Cherenkov light is focused onto pixel hybrid photon detectors~\cite{Alves:2008zz}, which provide fine spatial resolution and hence allow for the measurement of the Cherenkov angle. The data from RICH is processed using global likelihood approach~\cite{Forty:1998eqa}, by comparing various particle type hypotheses for each of the tracks. The PID information is then aggregated per charged track in the form of differences between log-likelihood values for a given particle type hypothesis and a pion hypothesis for that track. These differences are named \texttt{RichDLL*}, `\texttt{*}'~standing for \texttt{k} (kaon), \texttt{p} (proton), \texttt{mu} (muon), \texttt{e} (electron) and \texttt{bt} (below the threshold of emitting Cherenkov light); i.e. \texttt{RichDLLp} would, for example, stand for the log-likelihood difference between a proton and a pion hypothesis for a given track and therefore be used to distinguish between these two particle types.

\begin{figure}
    \centering
    \includegraphics[width=0.9\textwidth]{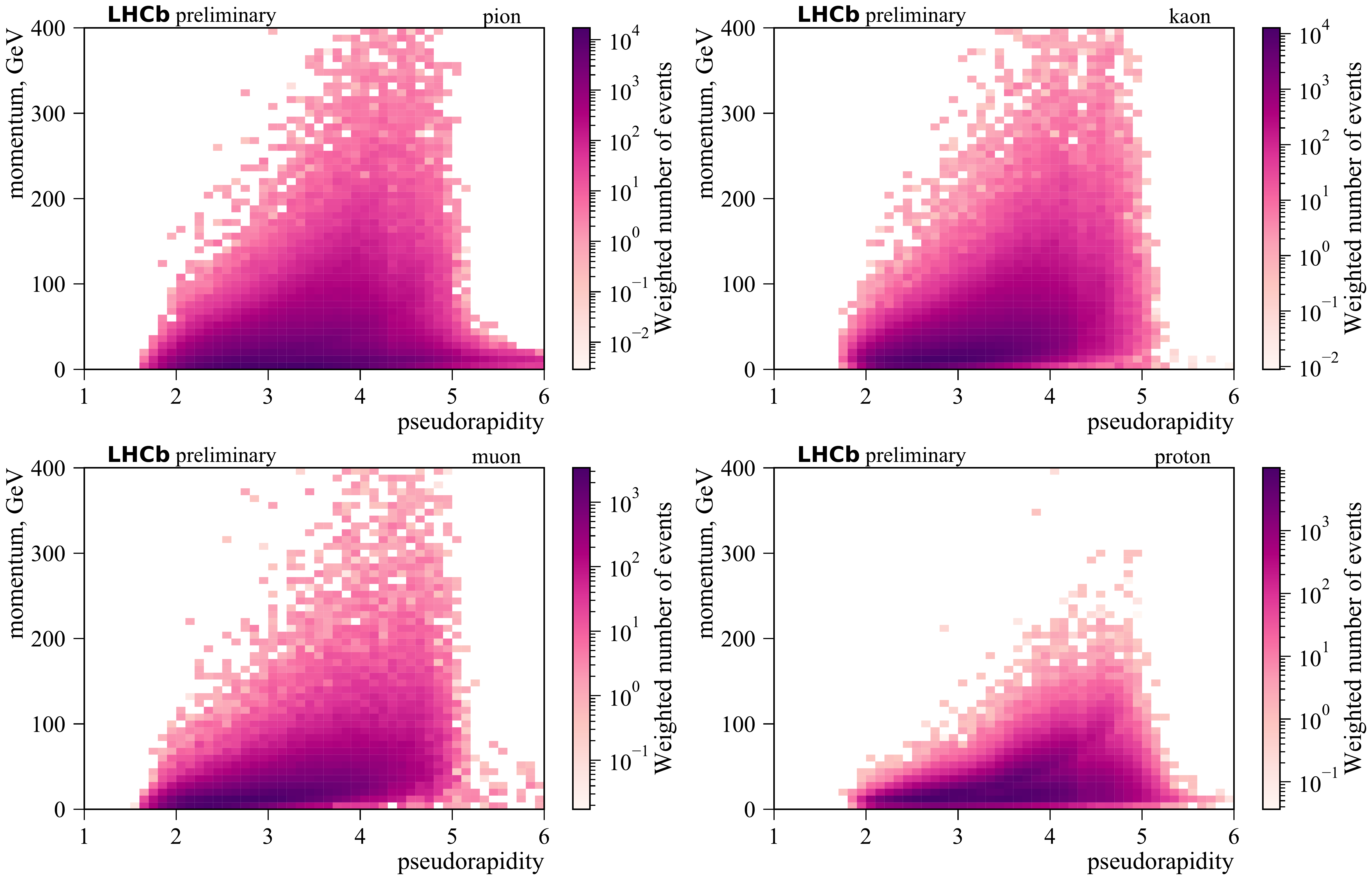}
    \caption{Momentum-pseudorapidity distributions for the calibration samples used for GAN training: pions~(top-left), kaons~(top-right), muons~(bottom-left) and protons~(bottom-right). The distributions are weighted using the technique described in text.}
    \label{fig:P-ETA}
\end{figure}

In this work, we train our model on data from several calibration samples collected in 2016~\cite{CalibrationSamples}. These are pure samples of charged tracks of different particle types that have been selected without the use of information from the PID subsystems. For each of the particle types we are training a separate independent instance of the model. We use muons from $\JpsiMuMu$ decays, kaons from $\DstPDzKpipi$ and $\DpsPhiPi$ decay channels, pions from $\DstPDzKpipi$ and $\KzsPiPm$ channels and protons from $\LzpPi$ decays\footnote{For each of the processes listed, both the process itself and its charge conjugate are implied.}. Figure~\ref{fig:P-ETA} shows the distributions of momentum and pseudorapidity for particles from calibration samples. These two quantities, along with the total number of reconstructed tracks in an event, form the input features to our neural networks, while \texttt{RichDLL*} values are the ones we are generating.

The calibration samples are not absolutely clean, and some amount of background is always present. The signal \texttt{RichDLL*} distributions are extracted from such data using the sPlot technique~\cite{Pivk:2004ty}. This method results in having non-unit sample weights such that weighted \texttt{RichDLL*} distributions are those of the signal component. These weights are applied to the loss functions during the training process. By design, they are allowed to be negative, which, as shown further, does not prevent the training process from convergence.

\section{Network architecture}

Both generator and discriminator are constructed with sets of fully connected layers of 128 neurons in each of the 10 hidden layers, with ReLU activation functions. The latent space size for the generator, i.e. the number of input noise neurons, is 64. Gaussian noise is used as the input for the generator. Cramer metric also allows to have non-unit output space size for the discriminator, which in our case is 256. The output layers of both generator and discriminator do not use activation functions. To further stabilize the training process, input and output data features are converted using Gaussian quantile transformer.

\section{Results}

\begin{figure}
    \centering
    \includegraphics[width=0.9\textwidth]{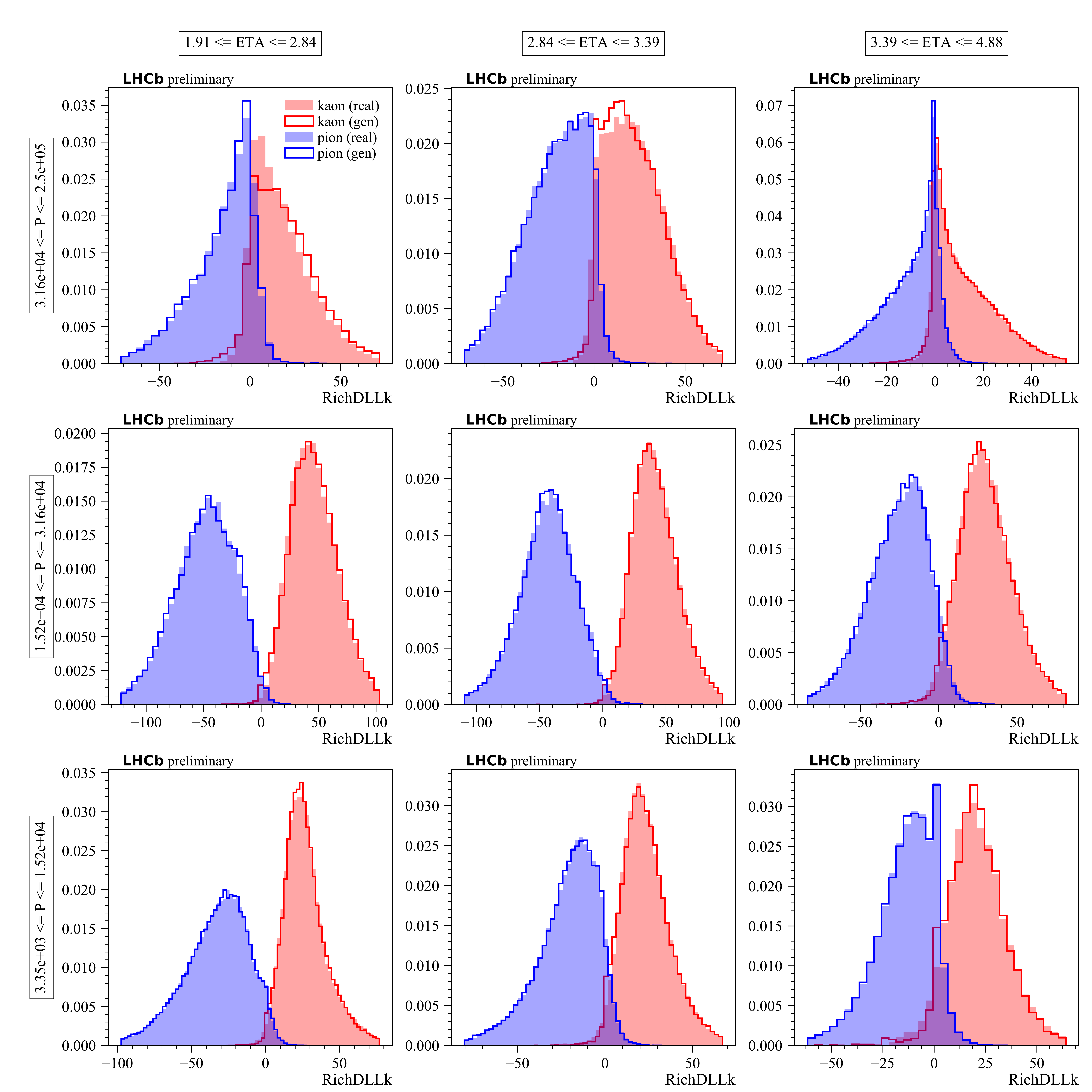}
    \caption{Weighted real data and generated distributions of \texttt{RichDLLk} for kaon and pion track candidates in bins of pseudorapidity (ETA) and momentum (P) over full phase-space.}
    \label{fig:RichDLLk}
\end{figure}

\begin{figure}
    \centering
    \includegraphics[width=0.7\textwidth]{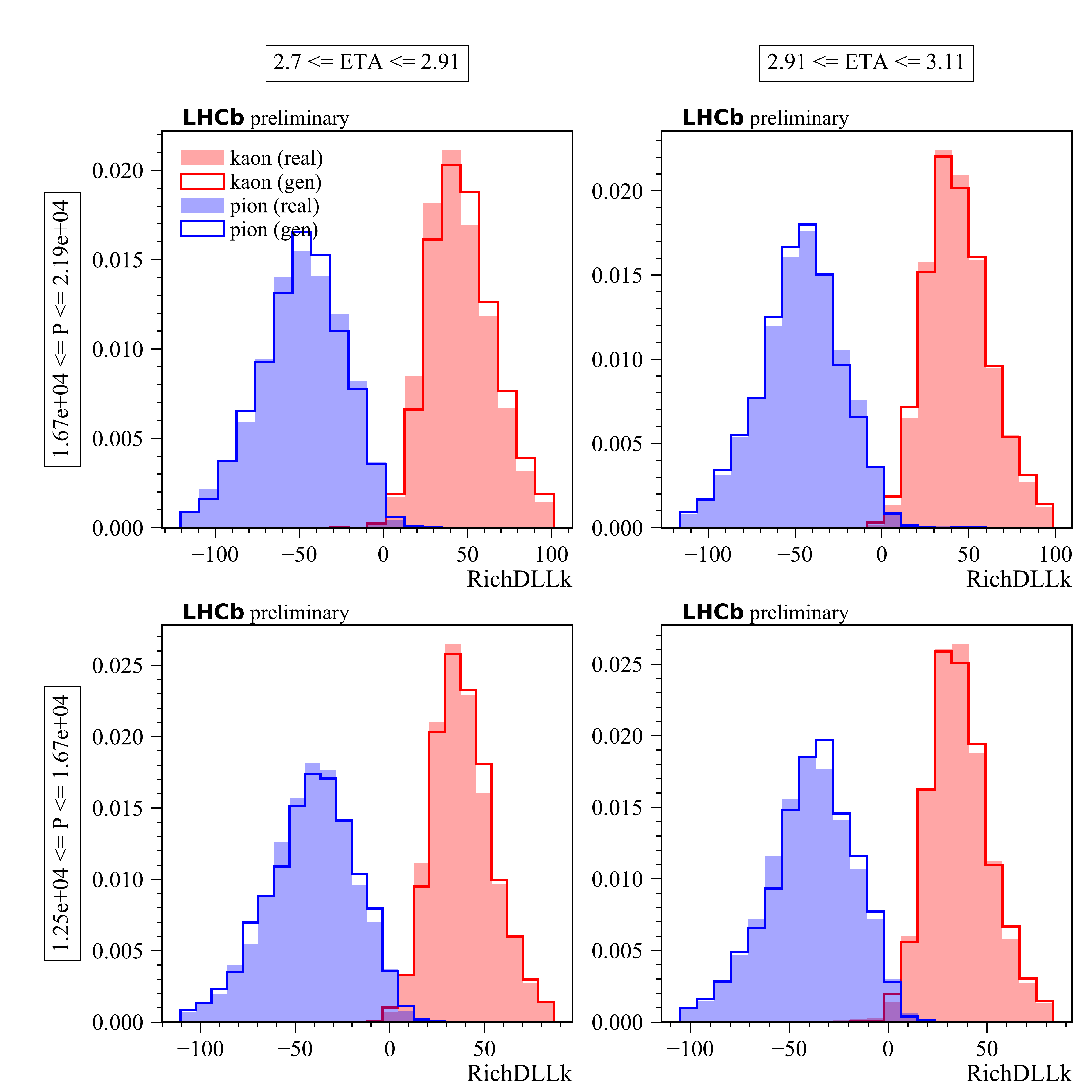}
    \caption{Weighted real data and generated distributions of \texttt{RichDLLk} for kaon and pion track candidates in bins of pseudorapidity (ETA) and momentum (P) in a well-populated phase-space region.}
    \label{fig:RichDLLkZoomed}
\end{figure}

Figure~\ref{fig:RichDLLk} shows comparison of weighted real data and generated distributions of \texttt{RichDLLk} for kaon and pion track candidates, in bins of momentum and pseudorapidity\footnote{The binning is only applied when plotting, while the model itself is trained with continuous input.}. The bin edges are selected to cover most of the phase space, such that each 1-dimensional bin has the same amount of kaons. Due to the correlations between momentum and pseudorapidity this condition cannot be satisfied in 2-dimensional bins. One can notice, that marginal bins are subject to smaller statistics and therefore slightly worse generation quality, while central bins show quite good correspondence to real data. Figure~\ref{fig:RichDLLkZoomed} shows similar distributions in a more narrow, but well-populated region of the phase-space.

\begin{figure}
    \centering
    \subfloat[kaons vs pions \label{fig:DAUCE:K}]{
        \includegraphics[width=0.31\textwidth]{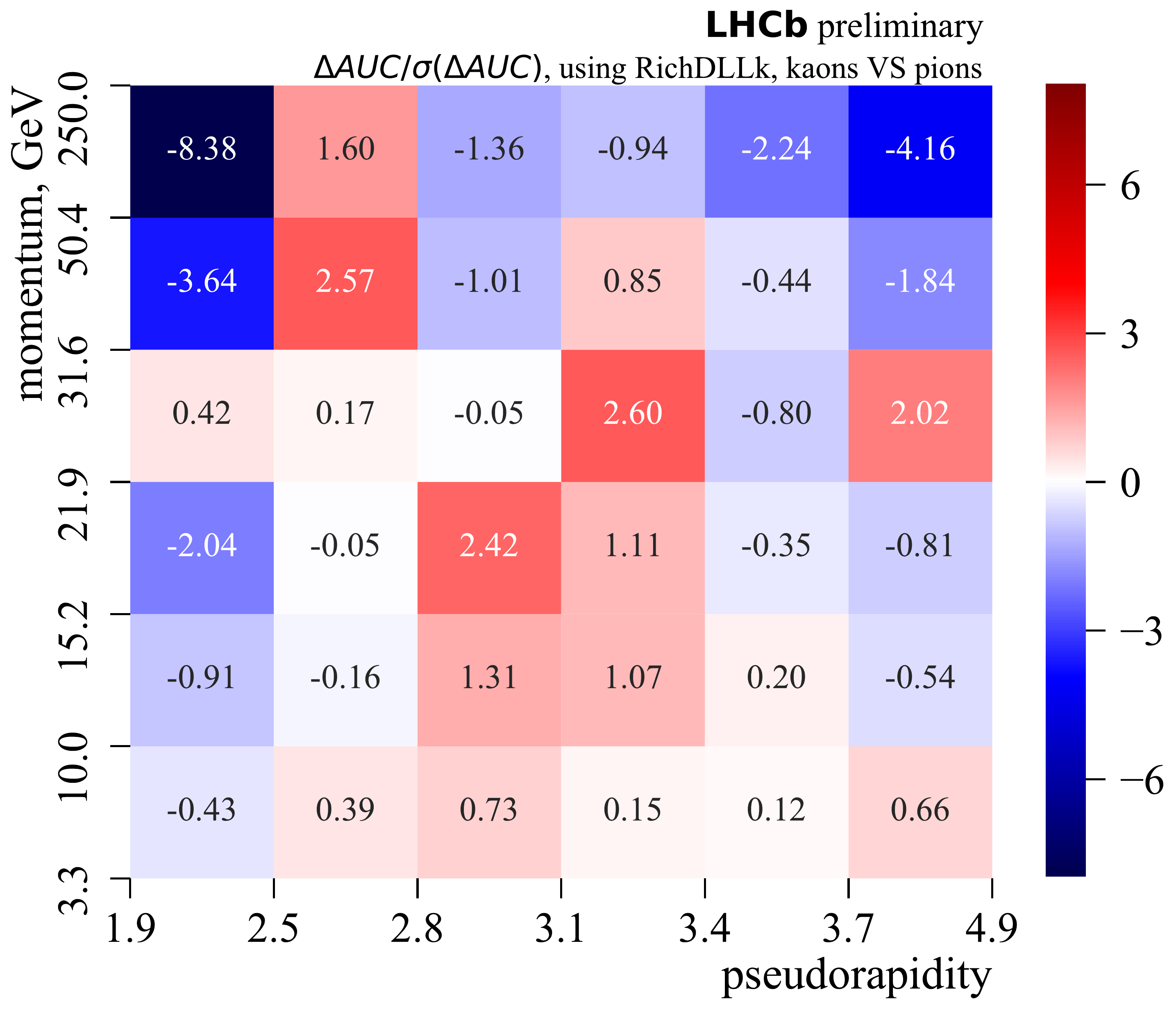}
    }
    \subfloat[muons vs pions \label{fig:DAUCE:MU}]{
        \includegraphics[width=0.31\textwidth]{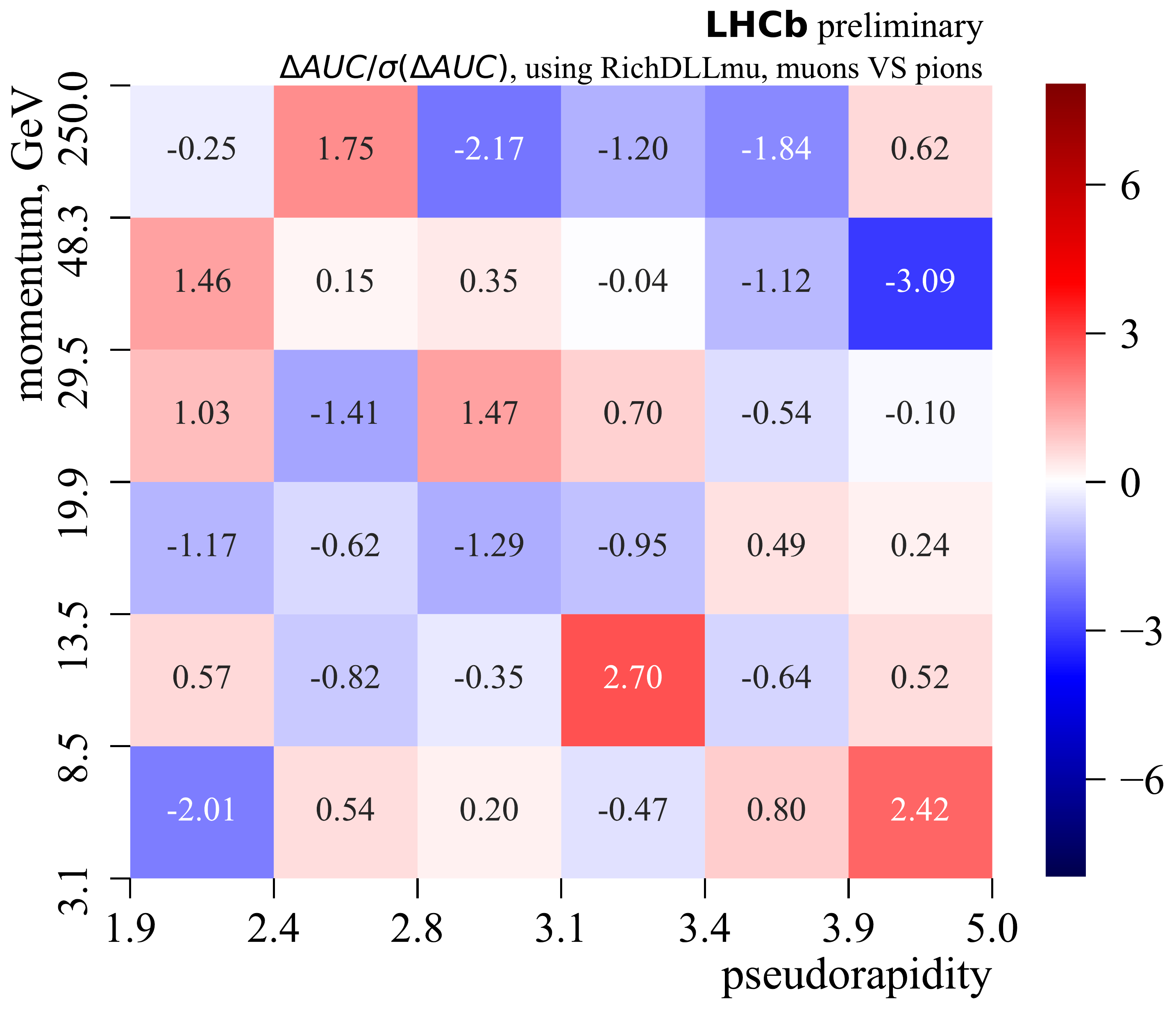}
    }
    \subfloat[protons vs pions \label{fig:DAUCE:P}]{
        \includegraphics[width=0.31\textwidth]{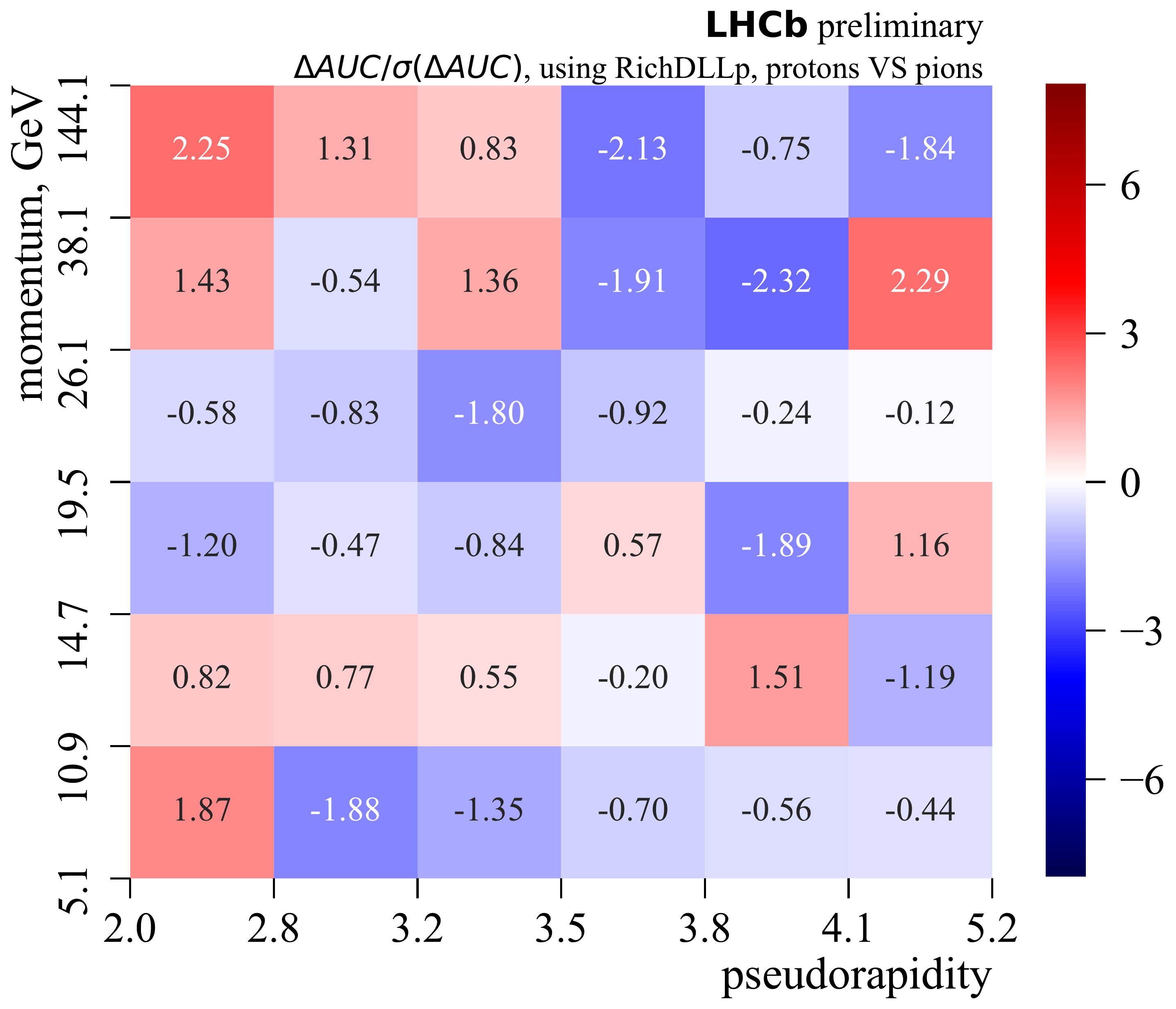}
    }
    \caption{Differences between real and generated sample areas under ROC-curves divided by uncertainties for discriminating kaons, muons and protons from pions, classifying with the \texttt{RichDLLk}, \texttt{RichDLLmu} and \texttt{RichDLLp} variables, respectively, in bins of momentum and pseudorapidity.}
    \label{fig:DAUCE}
\end{figure}

\begin{figure}
    \centering
    \subfloat[kaons vs pions \label{fig:DAUC:K}]{
        \includegraphics[width=0.3\textwidth]{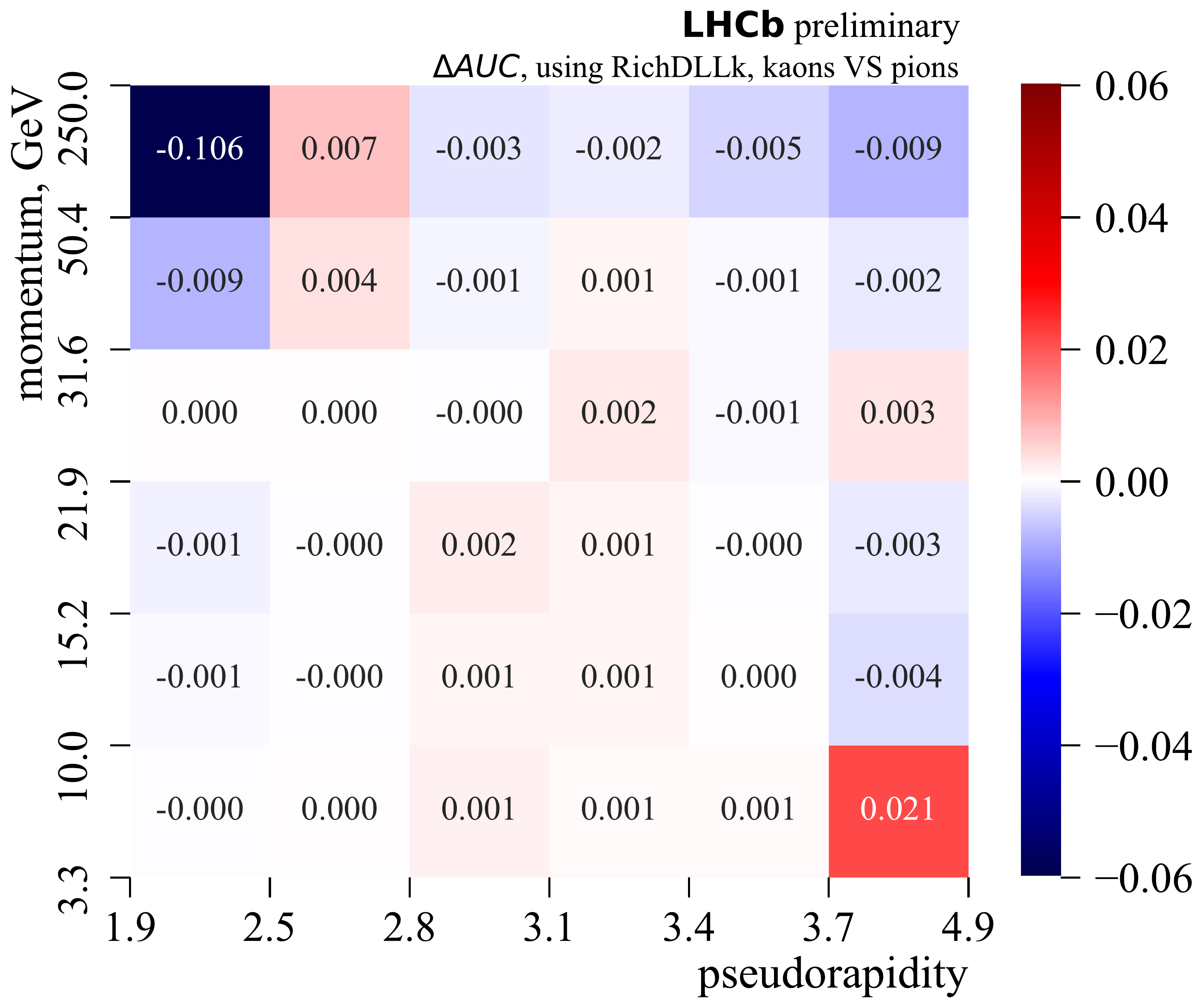}
    }
    \subfloat[muons vs pions \label{fig:DAUC:MU}]{
        \includegraphics[width=0.3\textwidth]{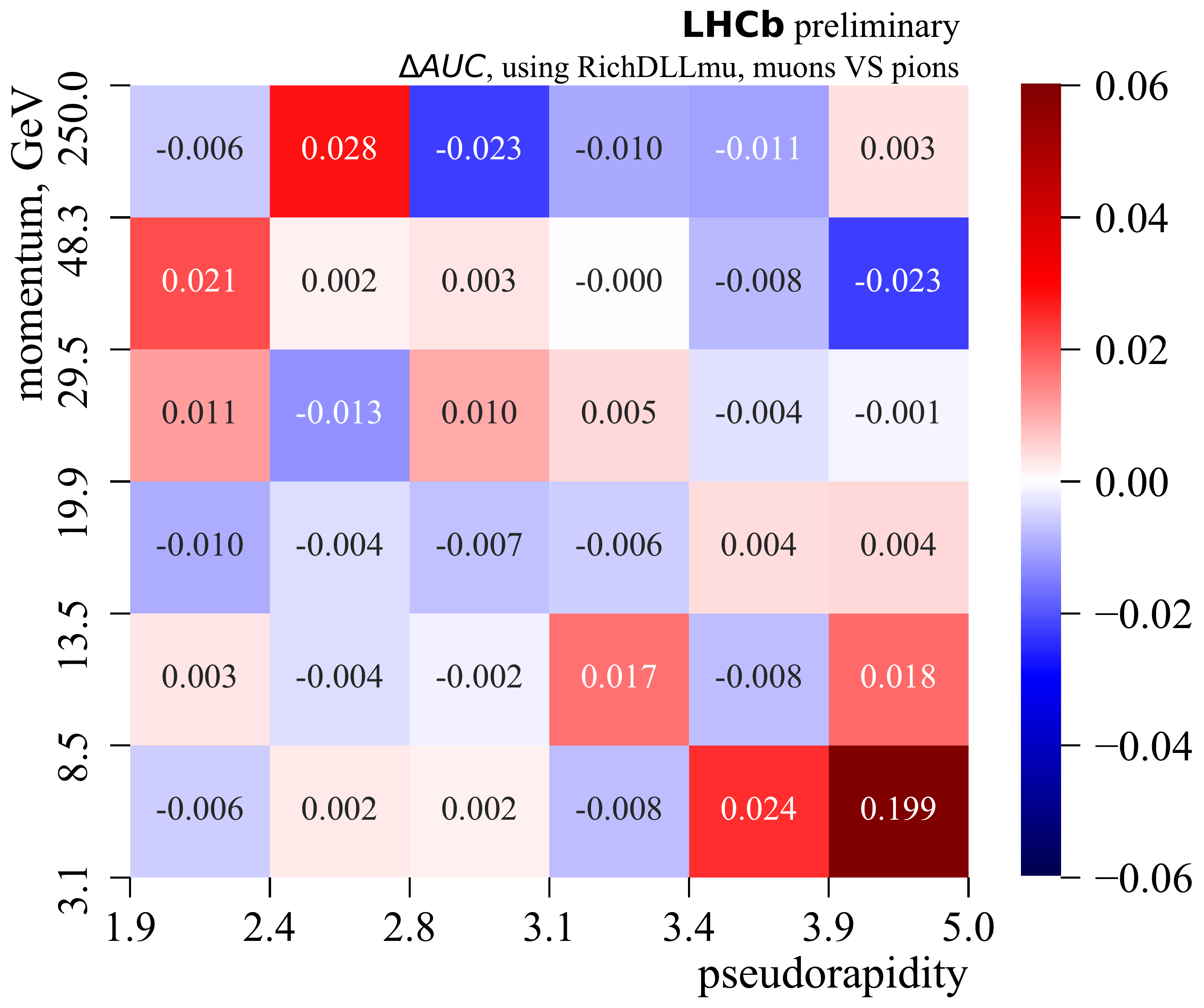}
    }
    \subfloat[protons vs pions \label{fig:DAUC:P}]{
        \includegraphics[width=0.3\textwidth]{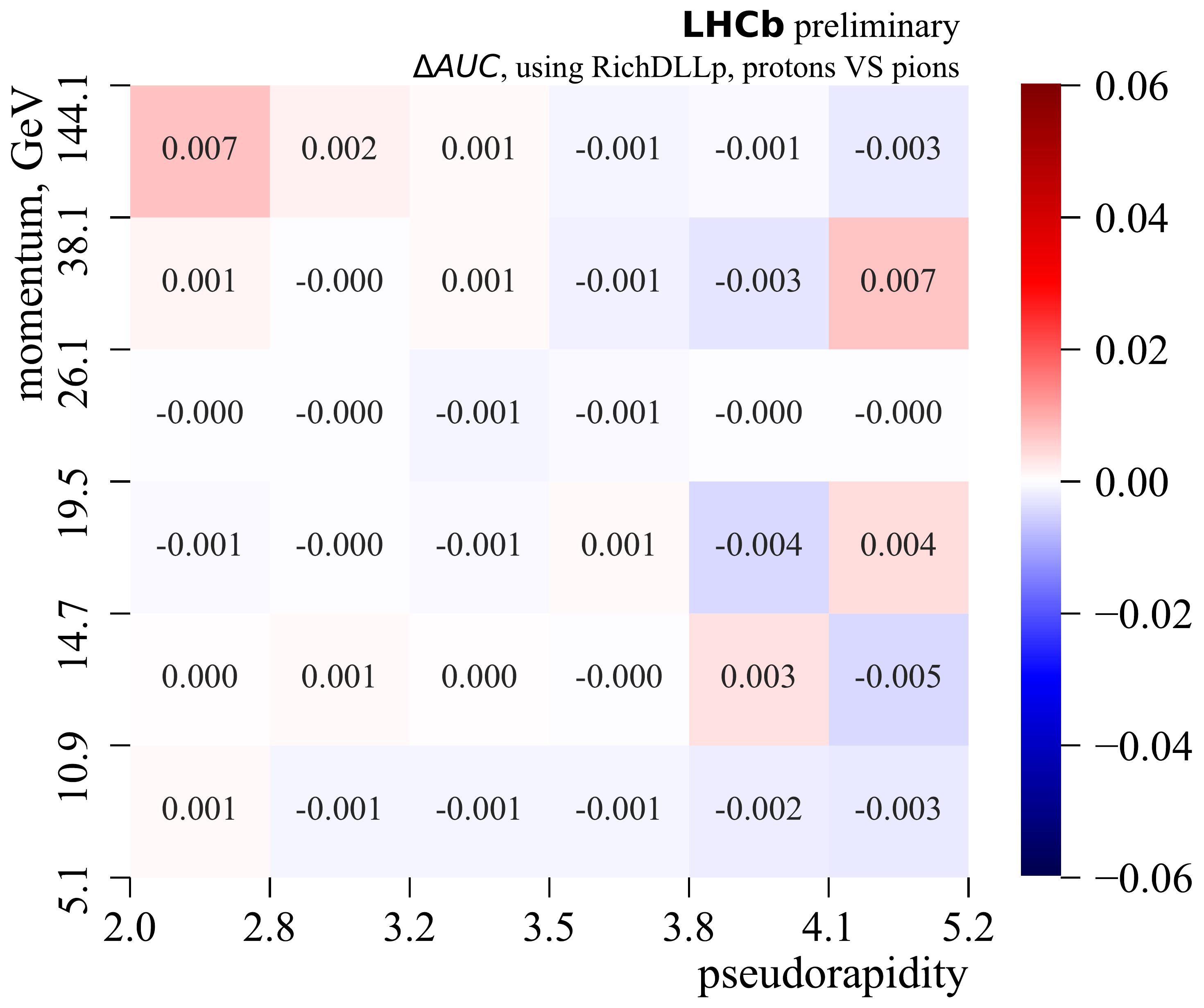}
    }
    \caption{Differences between real and generated sample areas under ROC-curves for discriminating kaons, muons and protons from pions, classifying with the \texttt{RichDLLk}, \texttt{RichDLLmu} and \texttt{RichDLLp} variables, respectively, in bins of momentum and pseudorapidity.}
    \label{fig:DAUC}
\end{figure}

In order to quantify the quality of the model in various regions of the phase space, area under the ROC curve (AUC) values were calculated in momentum-pseudorapidity bins for binary classification cases using both real data and generated variables. Figure~\ref{fig:DAUCE} shows differences between AUCs divided by uncertainties for real and generated samples for discriminating kaons, muons and protons from pions, classifying with the \texttt{RichDLLk}, \texttt{RichDLLmu} and \texttt{RichDLLp} variables, respectively, in bins of momentum and pseudorapidity. The uncertainty of the differences between AUC values was estimated using bootstrapping technique. One can see that most of the differences are not greater than few standard deviations, with no obviously biased regions, possibly with the exception of marginal bins that lack training statistics. Figure~\ref{fig:DAUC} shows the same differences not divided by the uncertainties, which demonstrates that most of them are of the order of $0.1-1\%$.

\section{Conclusion}

Fast simulation of the RICH detectors at LHCb can be achieved using generative models. In particular, GANs promise to be a good candidate for such generative approach. As training can be done on real data directly, there is no need for later tuning and corrections of the model, compared to the way regular accurate detector simulation algorithms are used.

The proposed model shows good approximation to the real data distributions with some imperfections in low-statistics regions. The training process converges in spite of the fact that sample weights are not strictly non-negative.

The demonstrated model quality evaluation does not provide the quantitative values for possible systematic uncertainties introduced by the model if used in a real physics analysis scenario. Evaluating such uncertainties is a subject for further investigations.

\ack
The research leading to these results has received funding from Russian Science Foundation under grant agreement n° 17-72-20127.

\section*{References}
\bibliographystyle{iopart-num}
\bibliography{references}
\end{document}

%% file: definitions.tex
\def \Jpsi                 {J/\psi}
\def \DstP                 {D^{+*}}
\def \Kzs                  {K^0_S}

\def \JpsiMuMu    {\Jpsi\rightarrow\mu^+\mu^-}
\def \DstPDzKpipi {\DstP\rightarrow D^0(K^-\pi^+)\pi^+}
\def \KzsPiPm     {\Kzs\rightarrow\pi^+\pi^-}
\def \DpsPhiPi    {D^+_s\rightarrow\phi(K^+K^-)\pi^+}
\def \LzpPi       {\Lambda^0\rightarrow p\pi^-}